# Theory of Optical Leaky-Wave Antenna Integrated in a Ring Resonator for Radiation Control

Caner Guclu, Ozdal Boyraz, and Filippo Capolino

*Abstract*—The integration of a leaky-wave antenna with a ring resonator is presented using analytical guided wave models. The device consists of a ring resonator fed by a directional coupler, where the ring resonator path includes a leaky-wave antenna segment. The resonator integration provides two main advantages: the high-quality factor ensures effective control of radiation intensity by controlling the resonance conditions and the efficient radiation from a leaky-wave antenna even when its length is much smaller than the propagation length of the leaky wave. We devise an analytical model of the guided wave propagation along a directional coupler and the ring resonator path including the antenna and non-radiating segments. The trade-offs regarding the quality factor of resonance and the antenna efficiency of such a design is reported in terms of the coupler parameters, leaky-wave constant and radiation length. Finally a CMOS-compatible OLWA design suitable for the ring resonator integration is designed where Silicon is utilized as the waveguide material for a possible electronic control of radiation intensity. The simulation results together with the analytical model show that slight variations in the leaky-wave's propagation constant, realized through excitation of excess carriers in Si domain, is sufficient to control the far-field radiation modulation with high extinction ratio.

## I. Introduction

OPTICAL antennas are proposed as a method of controlling optical radiation and also boosting light-matter interactions in the infrared and optical bands. Optical communication devices benefit highly from the well-established CMOS fabrication infrastructure and the integration with semiconductor technologies. To this aim, here we demonstrate a novel topology of a CMOS compatible optical leaky wave antenna (OLWA) that generates a controlled radiation beam.

Leaky-wave antennas, as a subcategory of guided wave antennas, have been studied extensively at the microwave frequency band [1], [2]. The radiation from a leaky-wave antenna occurs as a guided wave decays during its propagation, where such decay (exponential attenuation) is principally due to the radiated power leaking away during propagation along the waveguide. A so called slow-wave (hence guiding) structure can be transitioned into a leaky-wave antenna by introducing periodicity along the waveguide [3]. The radiation direction is mainly a function the leaky-wave phase constant, therefore control of this inherent characteristic grants the capability of frequency-controlled beam steering. Wide range of methods have been employed to design leaky-wave antennas, including metamaterial waveguides [4], frequency selective surfaces [5], multilayered substrates [6], [7]. For an up-to-date review of leaky-wave antennas, the reader is referred to [8] and to [9] for a review of the fundamental physics principles related to directivity using LW antennas. At optical and infrared frequency ranges, waveguides that host leaky-waves can be implemented by using linear arrays of polarizable particles. A theoretical frame of such a structure is reported for lossless penetrable spheres [10], and for a chain of plasmonic particles [11]. The leaky waves along plasmonic chains is theoreticaly studied in [12]–[14] and an extensive characterization of leaky and bound modes along a linear array of plasmonic spheres is presented in detail in [15]–[17]. Besides these examples of optical antennas based on plasmonic nanoparticles, plasmonic leaky-wave antennas [18], leaky-wave groove antennas [19] suitable for integration in optical chips and CMOS-compatible antenna topologies as in [20] constitute a strong option for controlling an efficient controllable off-chip radiation. OLWAs realized by patterning periodic small silicon inclusions in a CMOS-compatible dielectric waveguide were introduced [21] by some of the authors. These antennas combine very directive efficient radiation from low-loss dielectric waveguides at the infrared and optical frequencies with the advantages of utilizing semiconductor electronics. The control of excess carrier densities in silicon, thus its complex refractive index, was shown to provide limited but fundamental control of the radiation intensity and direction of an OLWA in [22], [23]. The radiation control capability of such an antenna was further boosted by integrating OLWAs with Fabry-Pérot resonator (FPR) formed by introducing reflectors along the propagation axis of the waveguide [15], the detailed theory and characterization of such an OLWA topology in 2-D model was presented in [24].

Manuscript received March 5, 2015. This work was supported in part by the National Science Foundation under NSF Award # ECCS-1028727.#

All the authors are with the University of California, Irvine, CA 92697 USA (corresponding author to provide phone: 949-824-2164; e-mail: cguclu@uci.edu, oboyraz@uci.edu, f.capolino@uci.edu).

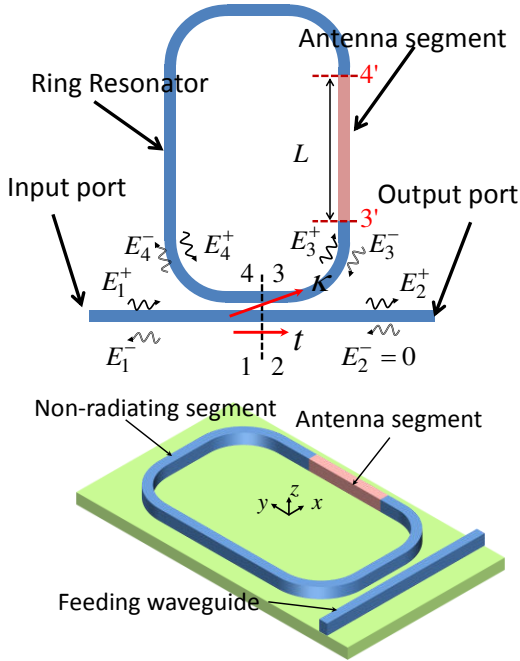

Fig. 1. The schematic of OLWA in RR and a 3-dimensional sketch representing the possible dielectric waveguide realization using a CMOS-compatible fabrication process.

In this work, an OLWA integrated with a dielectric waveguide ring resonator (RR) is introduced. The silicon waveguide-based RR including the antenna segment and its feed via a directional coupler can be realized with conventional CMOS fabrications process, and constitutes less demanding fabrication features than the OLWA in FPR [24]. The manuscript is organized as follows. Sec. II introduces a theoretical analytical model of the RR, directional coupler and the antenna segment based on a guided-wave schematic. The resonance conditions are developed, the antenna efficiency, the characterization of radiation from OLWA in RR and the dynamic radiation tunability are presented. In Sec. III, a simulated OLWA topology is investigated and related to the developed theoretical analytical model. We show the radiation control capabilities when a $10^{19}$ cm$^{-3}$ concentration of excess electron and holes are created in the silicon material of the antenna segment. The results prove the OLWA in RR as an effective wave of controlling efficiently LW radiation. They also prove the importance and effectiveness of the theory introduced in Sec. II in the design of resonator-integrated leaky-wave antennas at optical frequencies and any other frequency range.

## II. OLWA IN RING RESONATOR

The OLWA in ring resonator (OLWA in RR) is shown in Fig. 1 and it is composed of the radiating antenna segment (in pinkish orange color) and non-radiating guiding segments of the ring resonator (in blue color), the directional coupler made of the proximity region of the ring resonator and the feeding waveguide. In the following we will provide an analytical model explaining the working principles of OLWA in RR by using a schematic theoretical model, which is in principle valid in any frequency band from microwave to optical frequencies.

The resonator is shaped like a race track, and the bottom path of the resonator waveguide is brought in proximity with a feeding waveguide forming the directional coupler region. The radiation segment is depicted to lie on an arm of the resonator, and its orientation inside the resonator has no impact on the analytical model presented in the following.

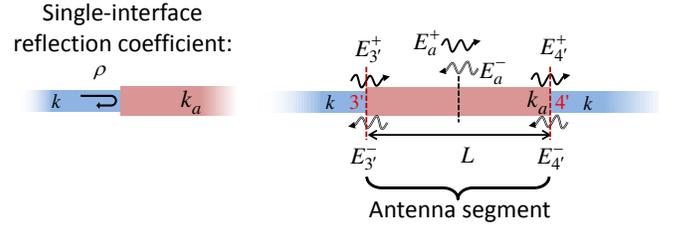

Fig. 2. (Left panel) The sketch depicting the definition the single-interface reflection coefficient. (Right panel) A close-up schematic view of the antenna segment.

In Fig. 1, the electric wave complex amplitudes are annotated with numeric indices denoting the ports of the directional coupler with the reference plane for the wave complex amplitudes denoted by the vertical dashed black line. Here the super script "+" denotes the E-field complex amplitude of the *direct* waves travelling from the input port to the output port along the feeding waveguide and circulating in the counterclockwise direction along the ring resonator. These would constitute the only waves if the mode mismatch at the two end of the antenna radiating segment was neglected. On the other hand, possible reflections at the discontinuities between the antenna and non-radiating segments along the resonator path (i.e., at sections 3' and 4'), would generate waves with clockwise propagation direction in the ring resonator. In turn, also waves propagating toward the input port along the feeding waveguide would be established and these are denoted by the super script "−". Moreover the network model illustrated in Fig. 2 (right panel) shows the electric field complex amplitudes $E_a^\pm$ propagating in the radiating antenna segment, that is referred to the center of the radiating segment for convenience, where + and − denote the counter clockwise and clockwise propagating waves, respectively. In the following $k$ and $k_a = \beta_a + i\alpha_a$ are the guided wavenumbers of the non-radiating and the radiating antenna waveguide segments, respectively (where the subscript "$a$" denotes the radiating *antenna* segment). Moreover, considering Fig. 2 (left panel), at the interface between the non-radiating and the radiating antenna segments, we denote by $\rho$ the electric field reflection coefficient referred to a wave impinging from the non-radiating waveguide. Note that this coefficient is defined between two semi-infinite waveguides, and the case of finite size of the antenna segment in Fig. 2 (right panel) is described using the scattering parameters as discussed regarding (2), later in the manuscript. Note that in an equivalent transmission line representation the reflection coefficient $\rho = (\bar{Z}_a - 1)/(\bar{Z}_a + 1)$ is represented in terms of the ratio $\bar{Z}_a$ of the characteristic impedances of the radiating and non-radiating segments in Fig. 2(left panel).

The radiation in the antenna segment takes place as a leaky-wave radiation. A leaky-wave is created by introducing spatial periodicity along the propagation direction. This leads to generating Floquet harmonics with wavenumbers $k_{a,q} = \beta_{a,q} + i\alpha_a$ where $\beta_{a,q} = \beta_a + q(2\pi/d)$ with $q = 0, \pm 1, \pm 2, ...$ and the $d$ is period (Fig. 3). In general, leaky-wave radiation from such structures are realized by tuning the $-1$ harmonic as a fast wave, i.e. $|\beta_{a,-1}| < k_h$ where $k_h$, assumed purely real, is the wavenumber in the host medium into which the radiation *leaks*. Therefore the radiation from the antenna segment is due to two leaky waves amplitudes $E_{a,-1}^{\pm}$, that represent the strengths of the $-1$ Floquet harmonics, whereas $E_a^{\pm}$ represents the fundamental harmonic (with $q = 0$). Assuming that the waveguide is slightly perturbed, the field in the waveguide can be approximated with the fundamental harmonic field. And the leaky-wave amplitudes $E_{a,-1}^{\pm}$ are such that $|E_{a,-1}^{\pm}| << |E_a^{\pm}|$, where $E_{a,-1}^{\pm}$ are linearly proportional to $E_a^{\pm}$. The accurate representation the field in the antenna segment and the assumptions applied above are discussed in detail in Appendix A.

Note that for simplicity we refer to electric field waves, in this paper, that are related to traveling power as $P = A|E|^2$ where the $A$ coefficient depends on the waveguide geometry, material and mode confinement properties and $E$ is the electric field complex amplitude of a particular wave shown in Fig. 1.

### A. Analytical Model

Utilizing the directional coupler model [25] where $\kappa$ and $t$ are the coupling and through coefficients, respectively, (assumed real in the following) we can construct four relations for +/− waves as

$$\begin{pmatrix} E_3^+ \\ E_2^+ \end{pmatrix} = \begin{pmatrix} i\kappa & t \\ t & i\kappa \end{pmatrix} \begin{pmatrix} E_1^+ \\ E_4^+ \end{pmatrix}, \quad \begin{pmatrix} E_1^- \\ E_4^- \end{pmatrix} = \begin{pmatrix} i\kappa & t \\ t & i\kappa \end{pmatrix} \begin{pmatrix} E_3^- \\ E_2^- \end{pmatrix}. \quad (1)$$

We assume that the output port of the feeding waveguide is terminated with perfect mode matching, thus we can assume $E_2^- = 0$ for simplicity.

The circulating field quantities at the antenna segment ends, ports 3' and 4' at the vertical red dashed lines in Fig. 2(right panel), are described by the electric field scattering parameters that can be associated also to the field at ports 3 and 4 as follows

$$\begin{pmatrix} E_{3'}^- \\ E_{4'}^+ \end{pmatrix} = \begin{pmatrix} s_{3'3'} & s_{3'4'} \\ s_{4'3'} & s_{4'4'} \end{pmatrix} \begin{pmatrix} E_{3'}^+ \\ E_{4'}^- \end{pmatrix},$$

$$\begin{pmatrix} E_3^- e^{-ikD_3} \\ E_4^+ e^{-ikD_4} \end{pmatrix} = \begin{pmatrix} s_{3'3'} & s_{3'4'} \\ s_{4'3'} & s_{4'4'} \end{pmatrix} \begin{pmatrix} E_3^+ e^{ikD_3} \\ E_4^- e^{ikD_4} \end{pmatrix}, \quad (2)$$

where $D_3$ ($D_4$) is the length of the non-radiating waveguide segment between the ports 3 and 3' (4 and 4'). Here the radiating antenna segment of length $L$ is a symmetric and reciprocal network, leading to $s_{3'3'} = s_{4'4'}$ and $s_{3'4'} = s_{4'3'}$ (see Appendix B). The sets of equations in (1) and (2) constitute a total of six equations with six unknown quantities ($E_1^-, E_2^+, E_3^+, E_3^-, E_4^+, E_4^-$) which can be solved in terms of the waveguide parameters, lengths of waveguide segments $L$, $D_3$ and $D_4$ and the incident electric field $E_1^+$. Moreover the network model illustrated in Fig. 2 (right panel) includes the electric field complex amplitude $E_a^{\pm}$, that is referred to the center of the radiating segment for convenience, where + and − denote the counter clockwise and clockwise propagating waves, respectively. Leaky-wave radiation modeled by equivalent aperture technique as in [24], [26], provides the radiation pattern through closed-form expressions in terms of the leaky wave wavenumber $k_{a,-1} = \beta_{a,-1} + i\alpha_a$ and the leaky-wave electric field $E_{a,-1}^{\pm} \propto E_a^{\pm}$. Assuming that the electric and magnetic field polarizations in the radiating antenna segment are mainly orthogonal to the waveguide axis the far-field pattern due to $E_a^{\pm}$ in the antenna segment depicted in Fig. 3 is the sum of the two beams created by the individual radiating waves given as

$$E_{FF} \propto \left[ E_a^+ \frac{\sin\left(\frac{L}{2}\chi^+\right)}{\chi^+} + E_a^- \frac{\sin\left(\frac{L}{2}\chi^-\right)}{\chi^-} \right] \cos\theta, \quad (3)$$

where $\chi^{\pm} = k_h \sin\theta \mp k_{a,-1}$, and $k_h$ is the wavenumber in the host medium into which the antenna radiation *leaks*. The two beams that are superposed in (3) exhibit maxima at angles $\theta \approx \pm \arcsin(\beta_{a,-1}/k_h)$, where $\beta_{a,-1} = \beta_a - 2\pi/d$. The characterization of the interference of such two beams is analyzed in depth in [24].

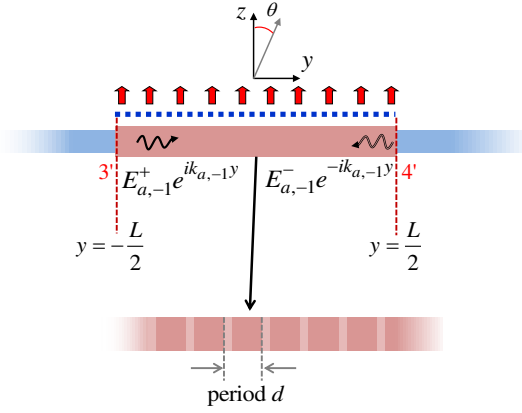

Fig. 3. The radiating antenna segment with the two radiating leaky wave harmonics propagating in opposite directions. Far-field radiation is found by applying equivalent aperture integration of the leaky wave fields along the blue dotted line. The antenna segment in our model is a periodically perturbed integrated waveguide with period $d$.

Therefore it is of importance to determine the leaky wave propagation wavenumbers and fields $E_a^+$ and $E_a^-$ that are derived in terms of the circulating waves' $E_3^+, E_3^-, E_4^+, E_4^-$. Upon algebraic manipulations, the radiating field complex amplitudes $E_a^\pm$ are found as

$$E_a^+ e^{-ik_a L/2} = E_3^+ e^{ikD_3} \frac{1}{1-\rho} + E_3^- e^{-ikD_3} \frac{-\rho}{1-\rho}, \quad (4)$$

$$E_a^- e^{ik_a L/2} = E_3^+ e^{ikD_3} \frac{-\rho}{1-\rho} + E_3^- e^{-ikD_3} \frac{1}{1-\rho}, \quad (5)$$

in terms of $E_3^\pm$. The equations (1) and (2) lead to the solutions of $E_3^\pm$ as

$$E_3^+ = E_1^+ \frac{i\kappa(1-t\Psi)}{(1-t\Psi)^2 - (t\Phi)^2}, \quad (6)$$

$$E_3^- = E_1^+ \frac{i\kappa s_{3'3'} e^{ik 2 D_3}}{(1-t\Psi)^2 - (t\Phi)^2}. \quad (7)$$

where we have defined $\Psi = s_{3'4'} e^{ik(D_3+D_4)}$ and $\Phi = s_{3'3'} e^{ik(D_3+D_4)}$ due to common occurrence. Thus, the leaky-wave electric field amplitudes and their ratio are derived as

$$\frac{E_a^+}{E_1^+} = \frac{e^{ik_a L/2} e^{ikD_3}}{1-\rho} i\kappa \frac{1-t\Psi - \rho s_{3'3'}}{(1-t\Psi)^2 - (t\Phi)^2}, \quad (8)$$

$$\frac{E_a^-}{E_1^+} = \frac{e^{-ik_a L/2} e^{ikD_3}}{1-\rho} i\kappa \frac{-\rho + \rho t\Psi + s_{3'3'}}{(1-t\Psi)^2 - (t\Phi)^2}. \quad (9)$$

$$\frac{E_a^+}{E_a^-} = e^{ik_a L} \frac{1-t\Psi - \rho s_{3'3'}}{-\rho + \rho t\Psi + s_{3'3'}}. \quad (10)$$

Note here that the ratio $E_a^+/E_a^-$ depends on the sum $(D_3 + D_4)$ and $L$ but not on $D_3$ or $D_4$ individually; therefore the magnitude ratio and the phase difference between the two radiating waves does not depend on where the antenna segment is positioned on the ring resonator. As clearly implied from (10), the phase difference $\angle(E_a^-/E_a^+)$ is determined by $L$. Together with the radiating field quantities, the next most important ones are the two-port device parameters. In order to assess the efficiency of the antenna, the reflection coefficient at the input port $\Gamma = s_{11}$ (since $E_2^- = 0$) and the transmission coefficient through the output port $T = s_{21}$ are, respectively, given by

$$\Gamma = \frac{E_1^-}{E_1^+} = \frac{-\kappa^2 s_{3'3'} e^{ik 2 D_3}}{(1-t\Psi)^2 - (t\Phi)^2}, \quad (11)$$

$$T = \frac{E_2^+}{E_1^+} = \frac{(1-t\Psi)\left[t - (t^2+\kappa^2)\Psi\right] - t(t^2+\kappa^2)\Phi^2}{(1-t\Psi)^2 - (t\Phi)^2}. \quad (12)$$

At resonance the common denominator in above quantities (8)-(12), $(1-t\Psi)^2 - (t\Phi)^2$, is to be minimized in order to enhance the fields in the antenna segment, $E_a^+$ and $E_a^-$.

Assuming that the directional coupler and the non-radiating segment of the ring resonator are lossless, in an input output description relative to ports 1 and 2, the power loss in the network model of OLWA in RR corresponds to the radiated power by the antenna segment. Accordingly we define the *total radiation efficiency* of the antenna as $\eta = P_{rad}/P_{inc}$, where $P_{rad} = P_{inc} - P_{refl} - P_{trans}$ is the power radiated by the OLWA, $P_{inc}$ and $P_{refl}$ are the incident and reflected powers at section 1 and $P_{trans}$ is the power transmitted after section 2. The total efficiency is then evaluated as

$$\eta = 1 - |\Gamma|^2 - |T|^2. \quad (13)$$

Ideally one would like that the total incident power $P_{inc}$ to be radiated, and thus the total efficiency $\eta = 1$.

### B. Small Reflectivity Approximation

While the derived equations constitute an accurate set of solutions for the antenna parameters, the assessment of the resonance condition and the response of OLWA in RR at resonance using these accurate equations is not straightforward. The characterization of OLWA in RR benefits much from a convenient approximation of assuming very small reflectivity as $|\rho|^2 \ll 0.25$. Under this assumption one has $\Psi^2 \ll \Phi^2$, therefore the common denominator in the expressions above is approximated as $(1-t\Psi)^2 - (t\Phi)^2 \approx (1-t\Psi)^2$ (for approximated $s_{3'3'}$ and $s_{3'4'}$, see Appendix B) Accordingly the radiating fields and two-port parameters of OLWA in RR are simplified to

$$\frac{E_a^+}{E_1^+} \approx \frac{e^{ik_aL/2}e^{ikD_3}}{1-\rho} \frac{i\kappa}{1-te^{ik_aL}e^{ik(D_3+D_4)}}, \quad (14)$$

$$\frac{E_a^-}{E_1^+} \approx \frac{e^{-ik_aL/2}e^{ikD_3}}{1-\rho} i\kappa\rho \frac{-1-i\frac{2\sin(k_aL)e^{ik_aL}}{1-te^{ik_aL}e^{ik(D_3+D_4)}}}{1-te^{ik_aL}e^{ik(D_3+D_4)}}, \quad (15)$$

It is apparent that in this case one has $E_a^- \propto \rho E_a^+$, and therefore the field in the radiating segment is dominated by the forward leaky wave, and the radiation pattern in (3) is provided by the first term mainly. The total reflection and transmission are approximated as

$$\Gamma \approx \frac{i\kappa^2 \rho e^{ik2D_3} \frac{2\sin(k_aL)e^{ik_aL}}{1-te^{ik_aL}e^{ik(D_3+D_4)}}}{1-te^{ik_aL}e^{ik(D_3+D_4)}}, \quad (16)$$

$$T \approx \frac{t - e^{ik_aL}e^{ik(D_3+D_4)}}{1-te^{ik_aL}e^{ik(D_3+D_4)}}, \quad (17)$$

where in the evaluation of $T$ we have also assumed that the directional coupler losses are negligible and thus $t^2 + \kappa^2 \approx 1$. The resonance condition is now straightforwardly determined by minimizing the absolute value of the denominator, $\left|1-te^{ik_aL}e^{ik(D_3+D_4)}\right|$. The resonance occurs when

$$e^{i\beta_aL}e^{ik(D_3+D_4)} = 1. \quad (18)$$

In other words, at resonance a circulating wave in the ring resonator accumulates a phase equal to an integer multiple of $2\pi$ and thus constructively superposes upon a round trip. At the resonance wavelength, the fields in the resonator are boosted. As $t$ gets closer to unity and the attenuation in the antenna segment decreases, the resonance gets sharper. Under a sufficiently sharp resonance, a very slight change in $\beta_a$ can cause the resonator to shift out of resonance, this way we can modulate the far-field intensity efficiently.

Given that the resonance condition is satisfied, i.e, when $e^{i\beta_aL}e^{ik(D_3+D_4)} = 1$, the approximated (for $|\rho|^2 \ll 0.25$) resonant values of the quantities in (14)-(17) take the form

$$\left.\frac{|E_a^+|}{|E_1^+|}\right|_R = \left|\frac{\kappa}{1-\rho}\frac{1}{1-te^{-\alpha_aL}}\right|, \quad (19)$$

$$\left.\frac{|E_a^-|}{|E_1^+|}\right|_R = \left|\frac{\kappa\rho}{1-\rho}\frac{-1-i\frac{2\sin(k_aL)e^{ik_aL}}{1-te^{-\alpha_aL}}}{1-te^{-\alpha_aL}}\right|, \quad (20)$$

$$|\Gamma|_R = \left|\frac{\kappa^2\rho\frac{2\sin(k_aL)e^{ik_aL}}{1-te^{-\alpha_aL}}}{1-te^{-\alpha_aL}}\right|, \quad (21)$$

$$|T|_R = \left|\frac{t-e^{-\alpha_aL}}{1-te^{-\alpha_aL}}\right|, \quad (22)$$

where the subscript "$R$" tags "resonance". Since the resonance can be sustained at any frequency by simply tuning the length of the non-radiating segment $(D_3+D_4)$, remarkably the expressions in (19)-(22) represent the *achievable* resonant values at any wavelength given that the resonance condition is satisfied by proper tuning of $(D_3+D_4)$. Two comments are to be made here, as observed from (20) and (21), the clockwise circulating radiating field and the reflection at the input port at resonance can be greatly suppressed by minimizing $\sin(k_aL)$ by setting $L = n\lambda_a/2$ with $n$ as an integer (i.e. setting the antenna segment length as a multiple of half a guided wavelength $\lambda_a = 2\pi/\beta_a$) thus antenna segment is also satisfying an additional resonance condition. Moreover the transmission coefficient $T$ can be minimized when $t \approx e^{-\alpha_aL}$ that makes the numerator in (22) vanish, i.e. when the critical coupling condition is satisfied as $\left|t-e^{-\alpha_aL}\right| \ll \left|1-te^{-\alpha_aL}\right|$ [25], [27], [28]. Therefore with the suppression of input reflection, $\Gamma$ one can increase the *total radiation efficiency* $\eta$ of the antenna. This is of extreme importance, since the efficiency of the leaky wave antenna in RR does not only depend on the antenna length, as in the case of a standard leaky wave antenna (not in a resonator) where the length is necessary to radiate all the power. The resonator effect can make even short leaky-wave antennas highly efficient.

### C. Illustrative Example

Here let us investigate an example case and assess how the OLWA in ring resonator can be used for modulation of radiation. The host medium is taken as vacuum with the vacuum wavenumber denoted by $k_0 = 2\pi/\lambda_0$ where $\lambda_0$ is the wavelength in vacuum. Let us take the wavenumber in non-radiating waveguide as $k = 1.55k_0$ whereas in the antenna segment the complex wavenumber is $k_a = (1.60+i0.001)k_0$. Note that since $\alpha_a/k_0 = 0.001$ is very small, there is a very low leaky wave attenuation in the radiating antenna segment corresponding to weak radiation for every single pass of the wave. However a low attenuation is a required condition for high RR quality factor and thus multiple wave passes through the radiating segment. The total length of non-radiating segments is taken as $D_3 + D_4 = 100\,\mu m$ (100 guided wavelengths $2\pi/k$ at the operation wavelength of 1550 nm), and the length of the antenna segment is taken as $L = 24.22\,\mu m$ (25 guided wavelengths $\lambda_a = 2\pi/\beta_a$ at $\lambda_0 = 1550\,nm$). In this example, the single-interface reflection coefficient (Fig. 2) is assumed to be $\rho = -0.0159$ and the directional coupler parameters are $\kappa = 0.31$ and $t = 0.95$. In Fig. 4 (top panel) the radiating field magnitudes [accurately evaluated through (8) and (9)] and their approximated resonant values [from (19) and (20)] versus free-space wavelength are reported, assuming a nominal $E_1^+ = 1$ V/m. For demonstration purposes the plotted wavelength range is chosen much larger than the free spectral range of the resonator (the distance between two consecutive peaks), so the periodic behavior of the resonant characteristics can be easily identified. The counterclockwise circulating radiating field $E_a^+$ as well as the clockwise circulating $E_a^-$ exhibit sharp peaks, though the latter is much weaker. Moreover at 1550 nm, $E_a^-$ is significantly suppressed as expected, because $\sin(k_a L)$ term in $E_a^-$ is minimized because we have chosen $L$ as an integer multiple of $\lambda_a/2$ at the resonance free-space wavelength of 1550nm (this also means that $|s_{3'3'}|$ is minimized). Moreover, the two dashed curves, $|E_a^-|_R$ and $|E_a^+|_R$, approximate very well the envelope of peaks, as achievable magnitude of $E_a^+$ and $E_a^-$ given that the RR resonates, i.e., by adjusting $(D_3 + D_4)$ so that (18) is satisfied at every wavelength. Especially the periodic behavior of the clockwise circulating $|E_a^-|_R$ is in very good agreement with the peak values of $|E_a^-|$ evaluated through exact expressions, proving that the approximated formulas (19) and (20) are suitable for estimating the exact field quantities at resonance wavelengths. In Fig. 4 (bottom panel), the magnitudes of the transmission and reflection coefficients (11), (12) versus wavelength are provided, together with their approximated resonant values (21), (22) denoted by dashed curves. At resonances $T$ exhibits sharp dips and $\Gamma$ exhibits sharp peaks. A particular case is observed at 1550 nm wavelength where the reflection is well suppressed, since the resonance of the antenna segment is also guaranteed by choosing $L$ as a multiple integer of $\lambda_a/2$, thus an increased efficiency at the resonance is sustained.

Next we will assume that the wavenumber in the radiating antenna segment is modified slightly as $k_a = (1.57 + i0.002)k_0$. Note that the imaginary part is increased significantly to observe the impact of attenuation along the ring resonator on the resonance. In reality such a change of $k_a$ may result in a slight variation of $\rho$, however here we keep $\rho$ the same for the sake of observing the effect of the change in $k_a$ only. The case with non-modified ($k_a = (1.60 + i0.001)k_0$ as in the previous case) and modified $k_a$ are denoted by the subscripts "I" and "II", respectively.

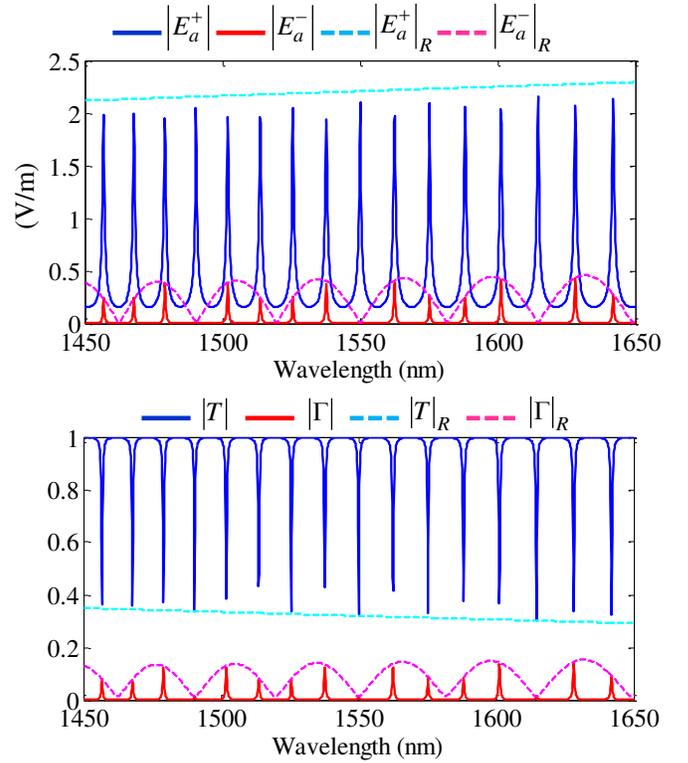

Fig. 4. (Top panel) The magnitude of two electric field waves in the radiating antenna segment, $|E_a^+|$ and $|E_a^-|$, and the resonant values $|E_a^+|_R$, $|E_a^-|_R$ versus wavelength. (Bottom panel) Transmission and reflection coefficients and their resonant values versus wavelength.

The total radiation efficiency (13) and the radiating field magnitude $|E_a^+|$ for cases I and II are reported in Fig. 5 top and bottom panels, respectively. The variation in $k_a$ results in the shift of resonance wavelengths as observed both by the efficiency and the radiating field curves. Moreover the peak values for case II are decreased compared to case I due to the increase in the attenuation constant $\alpha_a$, which decreases the resonance enhancement due to the denominator term $1 - te^{-\alpha_a L}$ and also resulting in loss of the critical coupling condition $t \approx e^{-\alpha_a L}$ previously discussed at the end of Sec. II-B. More strikingly, the radiating field magnitude $|E_a^+|$ at 1550 nm wavelength dropped significantly, mainly due to a change in the resonance frequency, therefore at a single frequency the variation of radiation is expected to be very significant. This phenomenon can be used to design tunable radiators, with large modulation depth.

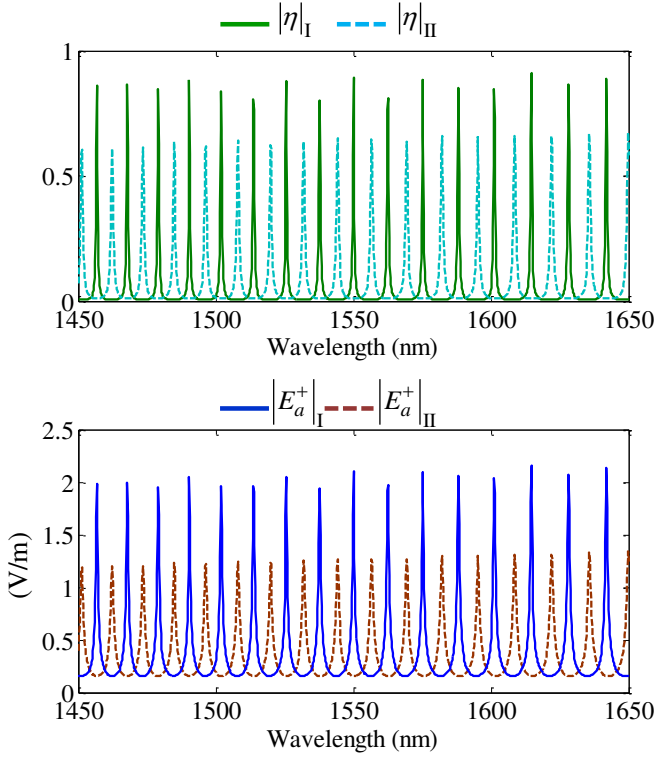

Fig. 5. (Top panel) Total radiation efficiency versus wavelength for cases I and II. (Bottom panel) Magnitude of counter clockwise propagating wave $|E_a^+|$ in the antenna segment versus wavelength for cases I and II.

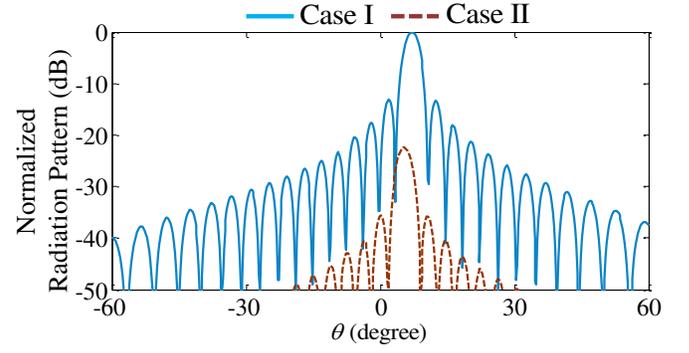

Fig. 6. Radiation pattern normalized by its maximum of Case I, versus observation angle $\theta$ with respect to $z$ as in Fig. 1.

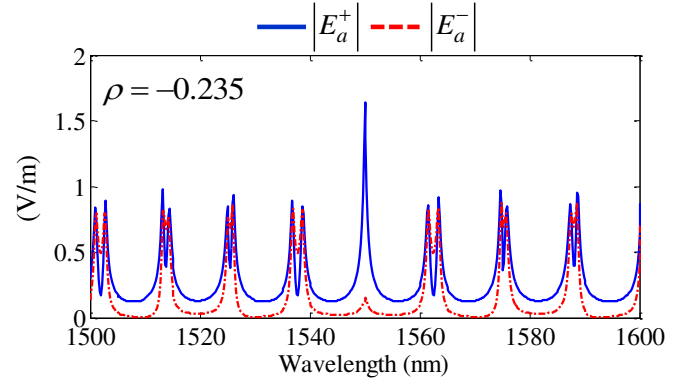

Fig. 7. Magnitude of electric field waves in the radiating antenna segment versus wavelength. In this case $\rho = -0.235$ is not negligible.

Now we aim to assess how the variation in the radiating field amplitudes $|E_a^+|$ translate to the far-field intensity. In Fig. 6 assuming a spatial periodicity $d = 1.05\,\mu\mathrm{m}$, we plot the radiation pattern evaluated via (3) for the two cases I and II with different $k_a$. Note that a change in $k_a$ as in Fig. 5 causes also a change in the leak-wave harmonic's wavenumber $k_{a,-1}$. The radiation patterns are normalized by the maximum intensity of Case I. We observe a main radiation direction at $\theta \approx 7°$ for Case I and at $\theta \approx 5.4°$ for Case II. There we observe a slight tilt in beam angle and in the intensity of $|E_a^+|$ is accompanied by a significant drop of 22 dB in the radiation intensity, as the resonator is kicked out of resonance due to the small variation of $k_a$. The far-field pattern is due to the counterclockwise circulating field $E_a^+$, as $E_a^-$ is negligibly smaller than $E_a^+$.

In Fig. 7 the magnitudes of $E_a^+$ and $E_a^-$ are reported for the same parameters as in Fig. 3 except for $\rho = -0.237$ which is significantly larger than previously taken, therefore the approximated formulas with $|\rho|^2 \ll 0.25$ are not strictly valid any more. In this case, the resonant condition for minimizing $(1 - t\Psi)^2 - (t\Phi)^2$ is more complicated. The resonant behavior is still observed, albeit with both *clockwise* and *counterclockwise* circulating waves in general reach *comparable* levels of enhancement at the resonances with multiple peaks occurring in an extremely narrow band. These cases can be exploited for beam tailoring since the total radiated beam can be tuned as a constructive interference of radiations due to $E_a^+$ and $E_a^-$, whose phase difference $\angle(E_a^+/E_a^-)$ can be adjusted by tuning $L$. The interesting resonance at 1550 nm wavelength stands out by exhibiting a clean single peak as in the previous cases. This special case occurs because at this wavelength the antenna segment has a resonant length $L = n\lambda_a/2$ and thus $|\sin(k_a L)| \ll 1$. This condition means that $|s_{3'3'}|$ is much smaller than unity, therefore $\Psi^2 \ll \Phi^2$ and the resonance condition reduces to minimization of $(1 - t\Psi)^2$ as in the case with $|\rho|^2 \ll 0.25$.

When $|s_{3'3'}| \approx 0$, the ratio $E_a^- / E_a^+$ in (9) tends to $-e^{\alpha_a L}/\rho$, allowing for tuning the radiating field ratio as a function of $\rho$.

## III. RADIATION CONTROL WITH CMOS-COMPATIBLE OLWA IN RR

The waveguide constituting the back bone of the practical example is made of Si, and deposited on 1-μm thick $SiO_2$ layer (with refractive index of 1.45) which is grown on a Si substrate. The Si waveguide has a uniform cross-section with a height of 0.8 μm and a width of 1 μm along the antenna non-radiating segment together with the feeding waveguide and the directional coupler region. Si chosen for its capability of refractive index modulation by controlling the free hole and electrons.

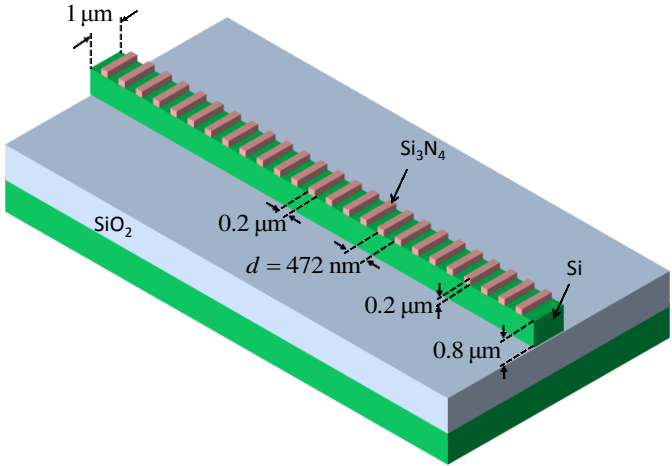

Fig. 8. Radiating segment of the OLWA embedded into the ring resonator. Periodic perturbations generate a leaky wave. Tunability of excess carriers in Silicon is used to control the radiated field.

A waveguide segment in the ring resonator is turned into the radiating antenna segment by utilizing periodically patterned 25 perturbations made of $Si_3N_4$ (with refractive index 2.1) on top of the Si waveguide. The perturbations has the same width as the waveguide (0.8 μm) and have a height of 0.2 μm. The perturbations are periodically patterned along the waveguide with a period of $d = 472$ nm and a duty cycle of 43% (i.e. a length of 200 nm). The ring resonator has $(D_3 + D_4) = 63.4$-μm long non-radiating and $L = 11.6$-μm long antenna segment. The coupler parameters are taken as $\kappa = 0.3$ and $t = 0.88$.

For demonstration of tunability the Si refractive index in the radiating antenna segment is assumed to be modified by controlling the excess hole and electron concentration therein. We assume the Si refractive index does not vary in other parts of the RR. The Si refractive index change is calculated using the Drude's formula [29]. Two states of excess carrier concentration are utilized in the following: The Si refractive index at 1550 nm wavelength for (i) State I with no excess carriers $N_{h,e} = 0$ is $n_{Si} = 3.48$, whereas for (ii) State II is $n_{Si} = 3.458 + i3.58 \times 10^{-3}$ when $N_{h,e} = 10^{19} \text{cm}^{-3}$. At 1550 nm free-space wavelength, the guided wavelength of the fundamental harmonic in the antenna segment $\lambda_a$ is 470.3 nm and 475.6 nm for State I and State II, respectively.

In the following simulations are carried out by using the finite element method solver HFSSS provided by ANSYS. For the radiating antenna segment, the simulated two-port scattering parameters are used in the schematic circuit model calculations as introduced in Sec. II. The injected power to the radiating OLWA segment is evaluated to be used in scaling the simulated radiation patterns. The simulated OLWA host an extremely slowly attenuating leaky-wave and a non-significant mismatch with the non-radiating waveguide segment. For both states I and II, $|s_{3'3'}|$ and $|s_{3'4'}|$ are nearly 0.08 and 0.95, respectively. In these cases the attenuation due to radiation dominates the losses of the resonator, hence the dissipative attenuation losses have negligible impact on the resonance behavior. With the simulated $|s_{3'4'}|$ and the coupler through coefficient $t = 0.88$, the antenna reaches nearly 80% radiation efficiency.

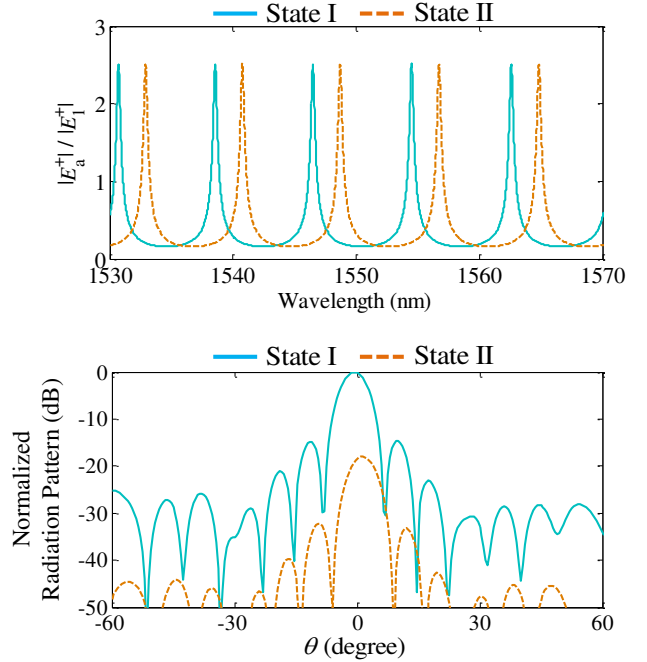

Fig. 9. (Top panel) Magnitude of electric field amplitude in the antenna segment (Fig. 2) for States I and II. (Bottom panel) Far-field radiation pattern for State I and State II at 1546 nm wavelength, normalized by the maximum of State I's gain. Changing state corresponds to a change in radiation intensity of 20dB.

The electric field amplitude $|E_a^+|$ inside the antenna segment is plotted in Fig. 9(top panel) normalized by the incident field $|E_1^+|$. There is an observable field variation between the States I and II, showing that at 1546 nm wavelength the field in the resonator can be efficiently brought to resonance or can be kicked out of resonance by switching between the states, mainly due to the variation in the guided wavelength in the antenna segment. In Fig. 9(bottom panel), we report the simulated radiation pattern of the OLWA, on the plane of propagation waveguide axis and the normal $z$-direction (i.e., in the so called "H-plane"). In the simulated radiation patterns, for State I compared to State II we observe a slight deviation in the radiation direction but a substantial drop in radiation intensity mainly because of ring resonator change of resonance condition. The change in the ration level is approximately 20dB and this property can be used to design modulators with deep modulation depth or switches,

## IV. CONCLUSION

The concept of a leaky wave antenna integrated in a ring resonator proved to be an effective way to control radiation intensity. We have shown that the integration of an OLWA into a ring resonator generates a large variation in the radiated intensity by slightly changing the refractive index of some component in the resonator. This open the way to a very effective way conceive electronically controlled switches with ;large extinction ration and modulators with large modulation depth. CMOS-compatibility of the proposed design and possible electronic device integration pave the way for efficient radiators based on semiconductor-based leaky wave antennas operating at infrared and optical frequencies. We also foresee a high speed of the control in case the waveguide is made of silicon nitride [30] and the perturbation with nanometer dimensions are made of silicon as in [31]. Importantly, the theory presented in this paper for the first time for a LW inside a ring resonator is applicable not only to optical antennas as discussed here, but to antennas at any frequency band. Indeed it can be applied also to leaky wave antennas designed in printed technology (e.g., microstrips) at millimeter and centimeter waves, for example.

## APPENDIX A

The guided mode in the radiating antenna segment (oriented along the $y$ direction) made of a periodically perturbed waveguide is represented as a weighted sum of Floquet harmonics, i.e., spatial harmonics, given by

$$E_{a,\text{mode}}(y) = \sum_{q=0,\pm1,\mp2,\ldots} E_{a,q} e^{ik_{a,q}y}. \qquad (23)$$

where $k_{a,q} = \beta_{a,q} + i\alpha_a$, and each harmonic has a phase propagation constant equal to $\beta_{a,q} = \beta_a + q2\pi/d$ where $d$ is the spatial period. Here $E_{a,\text{mode}}(y)$ represents the modal field profile along the waveguide at $y$, and it is a function of the unit cell material composition and geometry. The Floquet harmonic amplitudes are found by the projection integral over a unit cell length as

$$E_{a,q} = \frac{1}{2\pi} \int_{y=y_0}^{y=y_0+d} E_{a,\text{mode}}(y) e^{-ik_{a,q}y} dy. \qquad (24)$$

It is clearly seen that, the amplitude of the harmonics are linearly proportional to the field strength in the antenna segment.

Under the assumption of very slightly perturbed waveguide, the total field profile can be taken approximately equal to the fundamental mode's amplitude as $E_{a,\text{mode}}(y) \approx E_{a,0} e^{ik_a y}$ in the employed network model (in the manuscript, for convenience we drop the "0" subscript, which denotes the fundamental harmonic). It also follows that $|E_{a,-1}| \ll |E_{a,0}|$, though still maintaining the linear proportionality as $E_{a,-1} \propto E_{a,0}$. While negligible in the waveguide field, $E_{a,-1} e^{ik_{a,-1} y}$ is the leaky-wave harmonic amplitude, and it is the only term responsible for radiation. Therefore its wavenumber appear in the calculation of the far-field radiation (3). Note both leaky waves propagating in opposite directions in the radiating antenna segment share the same modal character description.

## APPENDIX B

The scattering parameters of the two-port antenna segment in between non-radiating waveguides, as a function of the LW wavenumber, LW radiation length and input reflection coefficient, are

$$s_{3'3'} = e^{ik_a L} \frac{i(-2\rho)\sin(k_a L)}{1-\rho^2 e^{ik_a 2L}}, \qquad (25)$$

$$s_{3'4'} = e^{ik_a L} \frac{1-\rho^2}{1-\rho^2 e^{ik_a 2L}}. \qquad (26)$$

Under the assumption of very low reflectivity $|\rho|^2 \ll 0.25$ utilized in Sec. II, these are approximated as

$$s_{3'3'} \approx i(-2\rho)\sin(k_a L)e^{ik_a L}, \qquad (27)$$

$$s_{3'4'} \approx e^{ik_a L}. \qquad (28)$$

## ACKNOWLEDGMENT

The authors are grateful to ANSYS, Inc., for providing HFSS. This work was supported by the National Science Foundation under NSF Award No. ECCS-1028727.## REFERENCES

[1] D. R. Jackson and A. A. Oliner, "Chapter 9 Leaky-Wave Antennas," in *Modern Antenna Handbook*, C. A. Balanis, Ed. John Wiley & Sons, Inc., 2008, pp. 325–367.
[2] A. A. Oliner and D. R. Jackson, "Chapter 12 Leaky-Wave Antennas," in *Antenna Engineering Handbook, Fourth Edition*, J. Volakis, Ed. McGraw-Hill Education, 2007, pp. 12–1.
[3] K. C. Gupta, "Narrow-beam antennas using an artificial dielectric medium with permittivity less than unity," *Electron. Lett.*, vol. 7, no. 1, pp. 16–18, Jan. 1971.
[4] C. Caloz, T. Itoh, and A. Rennings, "CRLH metamaterial leaky-wave and resonant antennas," *IEEE Antennas Propag. Mag.*, vol. 50, no. 5, pp. 25–39, Oct. 2008.
[5] S. A. Hosseini, F. Capolino, and F. De Flaviis, "Q-band single-layer planar Fabry-Perot cavity antenna with single integrated-feed," *Prog. Electromagn. Res. C*, vol. 52, pp. 135–144, 2014.
[6] D. R. Jackson, A. A. Oliner, and A. Ip, "Leaky-wave propagation and radiation for a narrow-beam multiple-layer

dielectric structure," *IEEE Trans. Antennas Propag.*, vol. 41, no. 3, pp. 344–348, Mar. 1993.

[7] H. Ostner, J. Detlerfsen, and D. R. Jackson, "Radiation from one-dimensional dielectric leaky-wave antennas," *IEEE Trans. Antennas Propag.*, vol. 43, no. 4, pp. 331–339, Apr. 1995.

[8] D. R. Jackson, C. Caloz, and T. Itoh, "Leaky-Wave Antennas," *Proc. IEEE*, vol. 100, no. 7, pp. 2194–2206, Jul. 2012.

[9] D. R. Jackson, P. Burghignoli, G. Lovat, F. Capolino, J. Chen, D. R. Wilton, and A. A. Oliner, "The Fundamental Physics of Directive Beaming at Microwave and Optical Frequencies and the Role of Leaky Waves," *Proc. IEEE*, vol. 99, no. 10, pp. 1780–1805, Oct. 2011.

[10] R. A. Shore and A. D. Yaghjian, "Travelling electromagnetic waves on linear periodic arrays of lossless spheres," *Electron. Lett.*, vol. 41, no. 10, pp. 578–580, May 2005.

[11] A. Alù and N. Engheta, "Theory of linear chains of metamaterial/plasmonic particles as subdiffraction optical nanotransmission lines," *Phys. Rev. B*, vol. 74, no. 20, p. 205436, Nov. 2006.

[12] X.-X. Liu and A. Alù, "Subwavelength leaky-wave optical nanoantennas: Directive radiation from linear arrays of plasmonic nanoparticles," *Phys. Rev. B*, vol. 82, no. 14, p. 144305, Oct. 2010.

[13] R. A. Shore and A. D. Yaghjian, "Complex waves on periodic arrays of lossy and lossless permeable spheres: 1. Theory," *Radio Sci.*, vol. 47, no. 2, p. RS2014, Apr. 2012.

[14] R. A. Shore and A. D. Yaghjian, "Complex waves on periodic arrays of lossy and lossless permeable spheres: 2. Numerical results," *Radio Sci.*, vol. 47, no. 2, p. RS2015, Apr. 2012.

[15] S. Campione, S. Steshenko, and F. Capolino, "Complex bound and leaky modes in chains of plasmonic nanospheres," *Opt. Express*, vol. 19, no. 19, pp. 18345–18363, Sep. 2011.

[16] A. L. Fructos, S. Campione, F. Capolino, and F. Mesa, "Characterization of complex plasmonic modes in two-dimensional periodic arrays of metal nanospheres," *J. Opt. Soc. Am. B*, vol. 28, no. 6, pp. 1446–1458, Jun. 2011.

[17] S. Campione and F. Capolino, "Waveguide and radiation applications of modes in linear chains of plasmonic nanospheres," in *Proceedings of 2013 URSI International Symposium on Electromagnetic Theory (EMTS)*, 2013, pp. 172–175.

[18] Y. Wang, A. S. Helmy, and G. V. Eleftheriades, "Ultra-wideband optical leaky-wave slot antennas," *Opt. Express*, vol. 19, no. 13, pp. 12392–12401, Jun. 2011.

[19] A. Polemi and S. Maci, "A leaky-wave groove antenna at optical frequency," *J. Appl. Phys.*, vol. 112, no. 7, p. 074320, Oct. 2012.

[20] K. Van Acoleyen, W. Bogaerts, J. Jágerská, N. Le Thomas, R. Houdré, and R. Baets, "Off-chip beam steering with a one-dimensional optical phased array on silicon-on-insulator," *Opt. Lett.*, vol. 34, no. 9, pp. 1477–1479, May 2009.

[21] Q. Song, S. Campione, O. Boyraz, and F. Capolino, "Silicon-based optical leaky wave antenna with narrow beam radiation," *Opt. Express*, vol. 19, no. 9, pp. 8735–8749, Apr. 2011.

[22] C. Guclu, S. Campione, O. Boyraz, and F. Capolino, "Enhancing radiation control of an optical leaky wave antenna in a resonator," in *Proc. SPIE 8497, Photonic Fiber and Crystal Devices: Advances in Materials and Innovations in Device Applications VI*, 2012, p. 84971J–84971J.

[23] S. Campione, C. Guclu, Q. Song, O. Boyraz, and F. Capolino, "An optical leaky wave antenna with Si perturbations inside a resonator for enhanced optical control of the radiation," *Opt. Express*, vol. 20, no. 19, pp. 21305–21317, Sep. 2012.

[24] C. Guclu, S. Campione, O. Boyraz, and F. Capolino, "Theory of a Directive Optical Leaky Wave Antenna Integrated into a Resonator and Enhancement of Radiation Control," *J. Light. Technol.*, vol. 32, no. 9, pp. 1741–1749, May 2014.

[25] *Integrated Ring Resonators*, vol. 127. Berlin, Heidelberg: Springer Berlin Heidelberg, 2007.

[26] W. L. Stutzman and G. A. Thiele, "Chapter 9 Aperture Antennas," in *Antenna Theory and Design 3rd ed.*, John Wiley & Sons, 2012, pp. 344–432.

[27] A. Yariv, "Universal relations for coupling of optical power between microresonators and dielectric waveguides," *Electron. Lett.*, vol. 36, no. 4, pp. 321–322, Feb. 2000.

[28] A. Yariv, "Critical coupling and its control in optical waveguide-ring resonator systems," *IEEE Photonics Technol. Lett.*, vol. 14, no. 4, pp. 483–485, Apr. 2002.

[29] S. Manipatruni, L. Chen, and M. Lipson, "Ultra high bandwidth WDM using silicon microring modulators," *Opt. Express*, vol. 18, no. 16, pp. 16858–16867, Aug. 2010.

[30] Y. Huang, Q. Zhao, L. Kamyab, A. Rostami, F. Capolino, and O. Boyraz, "Sub-micron silicon nitride waveguide fabrication using conventional optical lithography," *Opt. Express*, vol. 23, no. 5, pp. 6780–6786, Mar. 2015.

[31] Q. Zhao, C. Guclu, Y. Huang, S. Campione, F. Capolino, and O. Boyraz, "Experimental demonstration of directive Si3N4 optical leaky wave antennas with semiconductor perturbations at near infrared frequencies," 2015, vol. 9365, p. 93651K–93651K–10.